\documentclass[twocolumn,showpacs,preprintnumbers,amsmath,amssymb,superscriptaddress]{revtex4}

\usepackage{graphicx}
\usepackage{dcolumn}
\usepackage{bm}
\usepackage{color}

\begin{document}

\title{High-speed Jet Formation after Solid Object Impact}

\author{Stephan Gekle}
\affiliation{
Department of Applied Physics and J.M. Burgers
Centre for Fluid Dynamics, University of Twente, P.O. Box 217,
7500 AE Enschede, The Netherlands
}
\author{Jos\'e Manuel Gordillo}
\affiliation{
\'Area de Mec\'anica de Fluidos, Departamento de Ingener\'ia Aeroespacial y Mec\'anica de Fluidos, Universidad de Sevilla, Avda. de los Descubrimientos s/n 41092, Sevilla, Spain
}
\author{Devaraj van der Meer}
\author{Detlef Lohse}
\affiliation{
Department of Applied Physics and J.M. Burgers
Centre for Fluid Dynamics, University of Twente, P.O. Box 217,
7500 AE Enschede, The Netherlands
}

\date{\today}

\begin{abstract}
A circular disc impacting on a water surface creates a remarkably
vigorous jet. Upon impact an axisymmetric air cavity forms and
eventually pinches off in a single point halfway down the cavity.
Immediately after closure two fast sharp-pointed jets are observed
shooting up- and downwards from the closure location, which by
then has turned into a stagnation point surrounded by a locally
hyperbolic flow pattern. This flow, however, is {\it not} the
mechanism feeding the two jets. Using high-speed imaging and numerical
simulations we show that jetting is fed by the local flow around the base of the jet, which is forced by the colliding cavity walls. Based on this insight, we then show how the analytical description of a collapsing void (using a line of sinks along the axis of symmetry) can be continued beyond the time of pinch-off to obtain a quantitative model for jet formation which is in good agreement with the numerical and experimental data.
\end{abstract}


\maketitle


The most prominent phenomenon when a solid object hits a water
surface is the high-speed jet shooting upwards into the air. The
basic sequence of events leading to this jet has been studied since Worthington more than a century ago: After impact, the intruder creates an air-filled cavity in the liquid which due to hydrostatic pressure immediately starts to collapse, eventually leading to the pinch-off of a large bubble. Two very thin jets are ejected up- and downwards from the pinch-off point which constitutes a finite-time singularity intensively studied in recent time \cite{BurtonWaldrepTaborek_PRL_2005, GordilloEtAl_PRL_2005, KeimEtAl_PRL_2006,
BergmannEtAl_PRL_2006,GordilloPerezSaborid_JFM_2006, EggersEtAl_PRL_2007}. Such  singularities have been shown to lead to a hyperbolic flow pattern after collapse and thus to the formation of liquid jets \cite{LonguetHiggins_JFM_1983, ZeffEtAl_Nature_2000, DucheminEtAl_PhysFluids_2002}. 

As we show in the present work, however, the radial energy focussing towards the singular pinch-off point alone is not sufficient to explain the extreme thinness of jets observed after the impact of a solid object. Instead, this jet formation is shown to depend crucially on the kinetic energy contained in \textit{the entire collapsing wall of the cavity} even far above the pinch-off singularity.

This is in sharp contrast to jets observed in many other situations where narrow confining cavity walls are not present, e.~g. for bubbles bursting on a free surface or near a solid wall \cite{BoultonStoneBlake_JFM_1993, DucheminEtAl_PhysFluids_2002, BlakeEtAl_JFM_1993}, impact of liquid droplets \cite{OguzProsperetti_JFM_1990, Rein_FluidDynRes_1993, MortonRudmanLiow_PhysFluids_2000, BartoloJosserandBonn_PRL_2006, DengAnilkumarWang_JFM_2007}, wave focussing \cite{MacIntyre_JPhysChem_1968, ThoroddsenEtohTakehara_PhysFluids_2007} or jets induced by pressure waves \cite{AntkowiakEtAl_JFM_2007}. In all these cases jetting must thus be accomplished by a mechanism different from the one to be described in this paper. 

Our experimental setup consists of a circular disc with radius $R_0$ that is pulled through a water surface with velocity $V_0$ as described in \cite{BergmannEtAl_PRL_2006}. The velocity $V_0$ is kept constant throughout the whole process. Global and local Reynolds and Weber numbers are fairly large as shown in \cite{BergmannEtAl_PRL_2006} and therefore the only relevant control parameter is the Froude number, Fr$=V_0^2/R_0g$ with gravity $g$, which we choose to equal $5.1$ (for $R_0=2\mathrm{cm}$ and $V_0=1\mathrm{m/s}$).


Due to the large Reynolds numbers involved 
\footnote{After pinch-off one can additionally define $\mathrm{Re_{jet}}$ using the width of the jet at its base and the local free surface velocity. Also this Reynolds number is $O(10^3)$.}
and as during the highly transient process of jet formation only a negligible amount of vorticity is introduced into the system \cite{BoultonStoneBlake_JFM_1993} we treat the problem as inviscid and irrotational allowing for a potential flow description. We employ an axisymmetric boundary-integral technique which explicitly tracks the free surface. The topology change at pinch-off is implemented as follows: When the radial position of the node closest to the axis becomes smaller than the local node distance, the two neighboring nodes are shifted to the axis, conserving their vertical position and their potential. Continuing the simulation, these nodes eventually form the tip of the top and bottom jets. These numerical simulations have shown excellent agreement with experiment for different impact geometries \cite{BergmannEtAl_PRL_2006, GekleEtAl_PRL_2008} and we verified carefully that our results are independent of numerical parameters such as node density and time stepping. All simulations include surface tension but neglect the influence of air.

\begin{figure}
  \includegraphics[width=1\columnwidth]{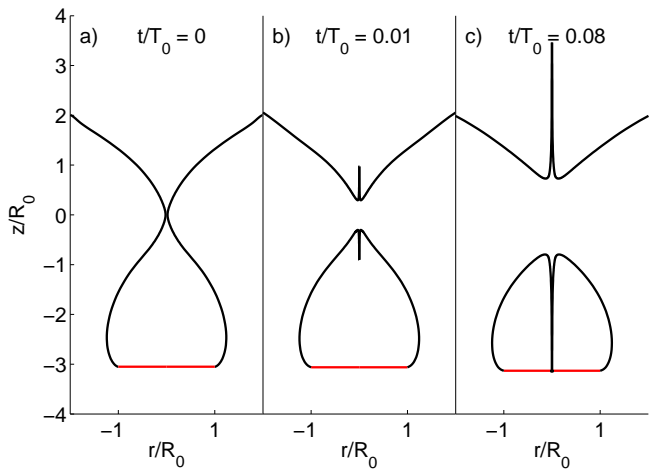}  
  \caption{The free surface shape (black) and the disc position (red) from the simulation at pinch-off (a), at an intermediate time with the growing up- and downward jets (b) and at the instant when the downward jet hits the disc (c).}\label{fig:surfaceProfiles}
\end{figure}

Figure~\ref{fig:surfaceProfiles} shows the pinch-off of the impact cavity and the subsequent formation of two thin jets. We use polar coordinates with $z=0$ at the pinch-off height and $t=0$ at the pinch-off moment. Velocity, length and time scales are normalized by $V_0$, $R_0$, and $T_0=R_0/V_0$, respectively. 

\begin{figure}
  \includegraphics[width=0.9\columnwidth]{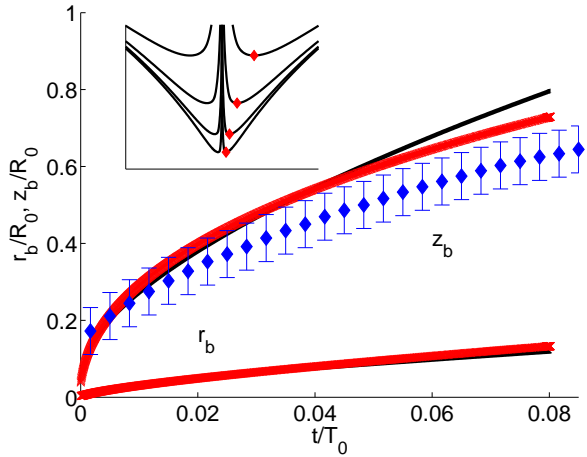}  
  \caption{The inset illustrates the position of the jet base (red diamond) at different instants of time. In contrast to other physical jetting situations the upward motion is much faster than the widening of the jet base. The main figure shows the upwards motion $z_b(t)$ of the jet base predicted by the analytical model (black line) which compares very favorably with the simulation (red crosses) and the experimental data (blue diamonds). (The slightly slower motion in the experiment can be attributed to an imperfect axisymmetry reducing the radial focussing effect.) The agreement between the model and the numerics is equally good for the widening of the base $r_b(t)$. Here, due to optical difficulties no experimental data are available.}\label{fig:baseDynamics}
\end{figure}

We set out to elucidate the precise mechanism which turns the horizontally collapsing cavity of Fig.~\ref{fig:surfaceProfiles} (a) into the thin vertical jets of Fig.~\ref{fig:surfaceProfiles} (b) and (c). For this we focus our attention on the dynamics of the upward jet base which we define as the local surface minimum illustrated in Fig.~\ref{fig:baseDynamics}. It is remarkable how the geometric confinement of the narrow cavity forces the jet to move upwards very fast while the widening of its base is restricted by the collapsing walls. We find that jet formation occurs on an extremely short time scale: the upwards jet grows above the initial quiescent surface in less than 1\% of the total time after impact. 

\begin{figure}
  \includegraphics[width=0.8\columnwidth]{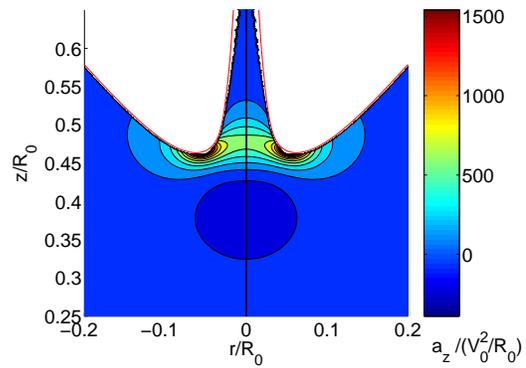}
  \caption{The vertical material acceleration $a_z=Dv_z/Dt$ at $t/T_0 = 0.028$ is confined to a small region around the jet base. The pinch-off location at (0,0) lies far too deep to influence the jetting process longer than in the first instants after pinch-off.}\label{fig:acceleration}
\end{figure}

These high speeds, however, are not due to a hyperbolic flow around the original pinch-off point as one could have expected based on suggested jetting mechanisms in other situations \cite{LonguetHiggins_JFM_1983, ZeffEtAl_Nature_2000, DucheminEtAl_PhysFluids_2002}. Figure~\ref{fig:acceleration} demonstrates how the fluid here is not accelerated upwards continuously from the pinch-off singularity but instead acquires its large vertical momentum in a small zone located around the jet base: Since each horizontal cross-section of the axisymmetric cavity wall will keep flowing radially inwards even after pinch-off, eventually it must collide on the axis in a similar way as the original pinch-off. This creates an upward and downward acceleration, of which the upward acceleration feeds the jet. The downward (negative) acceleration below the jet base can clearly be observed in Fig.~\ref{fig:acceleration}. It is thus essential to consider not only the singularity itself but the continuous collapse of the entire cavity wall in any kind of theoretical modelling.

Inspired by the above observations we now derive an analytical model for the jet formation as follows: As a first step the flow field of the collapsing cavity before pinch-off will be described by a line of sinks along the axis of symmetry \cite{GordilloPerezSaborid_JFM_2006, EggersEtAl_PRL_2007}. The strength of these sinks will be determined from the simulation {\it at pinch-off} and forms the only input quantity required by our model. Next, we will show how this picture naturally leads to a good description of the bulk flow after pinch-off. The line of sinks acquires a ''hole'' between the two jets and an additional point sink emerges near the jet bases. Finally, we will obtain two differential equations for the widening and upward motion of the jet base which are the two most relevant processes for jet formation. Secondary processes as jet breakup and the precise dynamics of the jet tip are not addressed in this work.

\begin{figure}
  \vspace{-1cm}
  \includegraphics[width=0.6\columnwidth, angle=270]{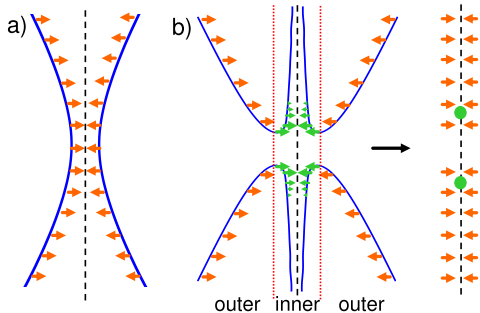}  
  \caption{(a) Sketch of the collapsing cavity being described by a distribution of sinks (orange arrows) on its free surface. (b) During jet formation the cavity collapse in the outer region remains unchanged (orange arrows) while around the jet base sinks accumulate (green arrows). This can be approximated as a line of sinks along the axis of symmetry plus a point sink (green dot). In the central region around the pinch-off point, a hole is formed: sinks are completely absent. For a detailed description see main text.}\label{fig:sinksIllu}
\end{figure}

As a starting point, Green's identity allows us to write the potential at any point $\mathbf{r}$ in the liquid bulk as an integral of sources and dipoles over the free surface:
\begin{equation}
4\pi\phi\left(\mathbf{r}\right) = \int_S dS' \mathbf{n'}\cdot\left[ \frac{1}{\left|\mathbf{r}-\mathbf{r}'\right|}\nabla'\phi - \phi\left(\mathbf{r'}\right)\nabla'\frac{1}{\left|\mathbf{r}-\mathbf{r}'\right|}\right]
\label{eqn:Green}
\end{equation}
with the integration taken over the free surface $S$ as illustrated in Fig.~\ref{fig:sinksIllu} (a) and (b). Since the dipole term decays quickly as $1/\left|\mathbf{r}-\mathbf{r}'\right|^2$, the source term (which decays only as $1/\left|\mathbf{r}-\mathbf{r}'\right|$) will be the only relevant contribution to the integral if the observation point is chosen sufficiently far from the free surface. As the cavity close to pinch-off becomes slender, $\partial \phi/\partial n \approx \dot{R}$ for a point $R$ on the free surface. Since the surface has no overhangs we write $dS = 2\pi R dz$. Approximating the radial distance as $r-r'\approx r$ turns Eq.~(\ref{eqn:Green}) into \cite{GordilloPerezSaborid_JFM_2006, EggersEtAl_PRL_2007}: 

\begin{equation}
2\phi(r,z) = \int \frac{q_{\mathrm{axis}}(z',t)}{\sqrt{r^2 + \left(z-z'\right)^2}}dz'.
\label{eqn:sinkIntegral}
\end{equation}
with a time- and height-dependent line distribution of sinks $q_\mathrm{axis}(z,t)$ along the axis of symmetry. Keeping in mind the extremely short time scale of jet formation as compared to the cavity collapse, we can assume the sink strength to remain \textit{constant in time} from the moment of pinch-off $t_c$ onwards, $q_c(z)=q_{\mathrm{axis}}(z,t_c)$.

During jet formation we divide the free surface into two regions separated by the jet base. The outer region contains the collapsing cavity until the jet base, while the inner region extends from the base inwards to the axis of symmetry as sketched in Fig.~\ref{fig:sinksIllu} (b). The principal fluid motion in the outer region remains identical to the collapsing cavity before pinch-off. High up in the jet, the motion will be vertically upwards with negligible radial velocity and thus will not contribute to the integral Eq.~(\ref{eqn:sinkIntegral}). As a free surface particle travels through the jet base and further up into the jet, it transitions from one flow regime to the other by decelerating its initial radial motion and turning it into vertical momentum. Thereby, its contribution to the integral (\ref{eqn:sinkIntegral}) decays to a negligible amount. This decay of the sink strength cannot happen instantaneously which leads to an accumulation (see Fig.~\ref{fig:sinksIllu}) of sinks around the jet base and a corresponding inward motion in that area. The length over which the sinks decay and accumulate must be proportional to the radius of the jet base which is the only relevant local length scale, $Cr_b$, with $C$ a constant of order one.

At this point it is good to stress that our model is not constructed to describe the liquid flow inside the jet itself but is valid only for the bulk flow outside the actual jet. This is sufficient for the explanation of jet formation which is the focus of the present paper. The sinks on the axis thus always remain outside of the liquid domain which they aim to describe. 

From an observation point located at $r\gg r_b$, the contribution of the sinks accumulating around the base can be regarded as a point sink of strength $Cr_bq_c(z_b)$ since $q_c(z)\approx q_c(z_b)$ along the length $Cr_b$. The point sink will be located a small distance above the base which is again proportional to the local length scale, i.e., $z_{\mathrm{sink}}=z_b + C_{\mathrm{sink}}\cdot r_b$ introducing a second constant $C_{\mathrm{sink}}$ of order unity.

Similarly, the most relevant contribution of the outer region will be that part of the line integral closest to the observation point $\mathbf{r}$. For an observation point at an altitude similar to or lower than the jet base, this is the region close to the jet base where again $q_c(z)\approx q_c(z_b)$. To allow analytical treatment of the integral resulting from Eq.~(\ref{eqn:sinkIntegral}), we can thus at any given time assume a sink strength being {\it constant in space} along the entire axis above the jet base. Through the motion of the jet base this sink strength depends implicitly on time $q_b(t)=q_c(z_b(t))$. 

Combining the approximations of the preceding paragraphs, we are now able to give an analytical expression derived from Eq.~(\ref{eqn:sinkIntegral}) for the potential at any point ($r$, $z$) as a function of the base position $r_b$ and $z_b$:
\begin{eqnarray}
2\phi = \underbrace{q_b\int_{-\infty}^{\infty} \frac{dz'}{\sqrt{r^2+(z-z')^2}}}_{\mathrm{collapsing\;cavity}} - \underbrace{q_b\int_{-z_b}^{z_b} \frac{dz'}{\sqrt{ r^2 + (z-z')^2}}}_{\mathrm{hole}}\nonumber\\
  + \underbrace{\frac{Cqr_b}{\sqrt{r^2 + (z-(z_b+C_{\mathrm{sink}}r_b))^2}}}_{\mathrm{point\;sink}}.
\label{eqn:potentialJet}
\end{eqnarray}
The initial sink distribution is obtained from the numerics by calculating $q_c(z)=-R\dot{R}$ along the surface just once \textit{at pinch-off}. It forms the \textit{only} input quantity required by our jetting model. Note that, as we are dealing with the upwards jet, the effect of the point sink for the downward jet can be neglected.

In order to derive the desired ODEs for $r_b(t)$ and $z_b(t)$ we apply the Bernoulli equation with zero pressure $\partial \phi / \partial t +\left|\nabla\phi\right|^2/2 = 0$ on the free surface. We then employ Eq.~(\ref{eqn:potentialJet}) to obtain the first differential equation involving $\dot{r}_b(t)$ and $\dot{z_b}(t)$. The second ODE results from the application of the kinematic boundary condition at the jet base $\partial \phi / \partial z = \partial z_b / \partial t$. This leads to a closed system of two ODEs. The calculations are straightforward but lengthy and are omitted here. With $C=4.55$ and $C_{\mathrm{sink}}=0.63$ the agreement with simulations and experiment is remarkable as demonstrated by Fig.~\ref{fig:baseDynamics}. We stress that the model requires as its only ingredient the sink strength distribution \textit{at pinch-off}. 

\begin{figure}
  \includegraphics[width=\columnwidth]{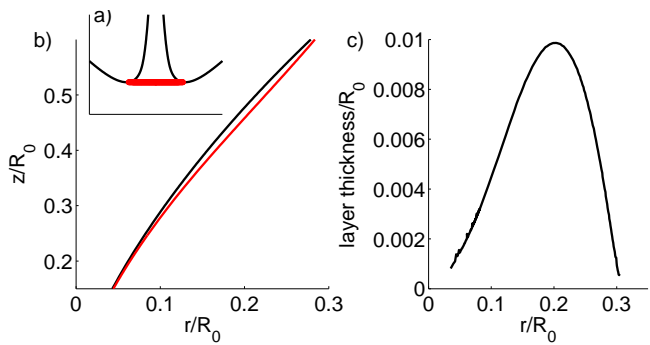}       
  \caption{A line of tracer particles is injected at the base of the jet at $t/T_0=0.08$ (a). After advecting them backwards in time to the moment of pinch-off they form the border of a thin surface layer containing the liquid that eventually ends up in the jet (b). A movie in which tracers are inserted at the base of the jet and advected backward in time is included as EPAPS document No. XXXXXX. The thickness of this layer is plotted in (c).}\label{fig:tracers}
\end{figure}

Finally, it is important to understand which region of the liquid bulk at pinch-off will eventually become ejected into the jet. This knowledge can be obtained from the numerical simulations by injecting a line of particles at the base of the jet, cf.~Fig.~\ref{fig:tracers} (a). Since the flow field is known for all times previous to particle injection, the tracers can be followed backwards to their origin at $t=0$. The line of tracers injected at the final instant will yield the outer boundary of the fluid layer that, together with the free surface, delimitates the fluid volume from which the jet originates. While the radial extent of the fluid layer depicted in Fig.~\ref{fig:tracers} (c) is of the order of the disc radius, its maximum thickness is only about $0.01\cdot R_0$. Thus far, a similar surface layer has only been observed when jetting is directly caused by surface waves \cite{MacIntyre_JPhysChem_1968}. In the present case, the thinness of the layer is even more remarkable as it does not arise from a surface phenomenon but from the collapsing motion of the entire bulk liquid.  

In conclusion, we have studied in detail the mechanism responsible for the
formation of high-speed Worthington jets after the impact of solid
objects on a liquid surface. We showed that the liquid forming the jet originates from a thin layer straddling the surface of the impact cavity. Our main finding, nevertheless, is the vital importance of the radial energy focussing \textit{along the entire wall} of this cavity. In contrast to other situations \cite{LonguetHiggins_JFM_1983, ZeffEtAl_Nature_2000, DucheminEtAl_PhysFluids_2002}, the hyperbolic flow around the singular pinch-off point turned out to be \textit{not} the relevant mechanism behind jet formation. Instead, our case seems more reminiscent of the violent jets observed during the explosion of lined cavities \cite{BirkhoffEtAl_JApplPhys_1948}. We proposed an analytical model which is in very good quantitative agreement with experimental data and numerical simulations. The only ingredients to the model are two constants of order one and a sink distribution $q_c(z)$ describing the collapsing cavity \textit{at pinch-off}. 

We expect that the present mechanism is also responsible for the very thin jets ejected after the impact of water droplets on a liquid pool \cite{Rein_FluidDynRes_1993} in a parameter range where a small cylindrical cavity at the bottom of the crater collapses in a very similar fashion as the impact cavity described in this work. In the future, our model of jet formation can serve as the base for predicting the shape and the velocity of the jet itself.

\vspace{-0.3cm} 

\begin{acknowledgments}
We thank A.~Prosperetti for discussions. This work is part of the program of the Stichting FOM, which is financially supported by NWO. JMG thanks the financial support of the Spanish Ministry of Education under projects DPI2005-08654-C04-02 and DPI2008-06624-C03-01.
\end{acknowledgments}

\end{document}